\documentclass[fonts]{icst}

\usepackage{moreverb}

\usepackage[breaklinks,colorlinks,bookmarksopen,bookmarksnumbered,linkcolor=ICSTblue,citecolor=blue,urlcolor=ICSTblue]{hyperref}
\usepackage{doi}

\newcommand\BibTeX{{\rmfamily B\kern-.05em \textsc{i\kern-.025em b}\kern-.08em
T\kern-.1667em\lower.7ex\hbox{E}\kern-.125emX}}

\def\volumeyear{2014}

\journalname{Collaborative Computing}
\articletype{Research Article}
 \def\volumeyear{XXXX}
 \def\volumenumber{XX}
  \def\issuemonth{XX-XX}
  \def\issuenumber{X}
  \def\articleid{XX}
  \def\copyrightholder{Simon\textit{ et al.}, licensed to ICST}
  \copyrightnote{This is an open access article distributed under the terms of the Creative Commons Attribution license (\url{http://creativecommons.org/licenses/by/3.0/}), which permits unlimited use, distribution and reproduction in any medium so long as the original work is properly cited.}
  \def\articledoi{10.4108/XX.X.X.XX}
\received{XXXX}
  \accepted{XXXX}
  \published{XXXX}

\begin{document}

\runningheads{J. Simon, P. Schmidt, V. Pammer-Schindler}{An Energy Efficient Implementation of DiffSync}

\title{Analysis of Differential Synchronisation's Energy Consumption on Mobile Devices}

\author{J\"org Simon\affil{1}\fnoteref{1}, Peter Schmidt\affil{2}, Viktoria Pammer-Schindler\affil{3}}

\address{
\affilnum{1}Know-Center GmbH, Inffeldgasse 13/6, 8010 Graz, Austria\\
\affilnum{2}Mendeley Ltd, White Bear Yard 144a Clerkenwell Road, London, UK\\
\affilnum{3}Knowledge Technologies Institute, Graz University of Technology, Inffeldgasse 13/5, 8010 Graz, Austria}

\abstract{
Synchronisation algorithms are central to collaborative editing software. As collaboration is increasingly mediated by mobile devices, the energy efficiency for such algorithms is interest to a wide community of application developers. In this paper we explore the differential synchronisation (diffsync) algorithm with respect to energy consumption on mobile devices. Discussions within this paper are based on real usage data of PDF annotations via the Mendeley iOS app, which requires realtime synchronisation.
\\
We identify three areas for optimising diffsync: a.) Empty cycles in which no changes need to be processed b.) tail energy by adapting cycle intervals and c.) computational complexity. Following these considerations, we propose a push-based diffsync strategy in which synchronisation cycles are triggered when a device connects to the network or when a device is notified of changes. }



\fnotetext[1]{Corresponding author email: \email{jsimon@know-center.at}}

\maketitle

\newcommand\tabhead[1]{\small\textbf{#1}}


\section{Introduction}
Enabling synchronous read and write access to shared items is one of
the key functionalities in collaborative software (e.g., svn), games
(e.g, Age Of Empires~\cite{Terrano_Bettner_2001}) or documents (e.g.,
Google Docs). In the wake of increased mobile broadband access
distributed collaboration is increasingly mediated by mobile
devices. Application developers should, therefore, consider energy
consumption in the use of synchronisation algorithms.
\\
In this paper we explore the energy consumption of the differential
synchronisation algorithm (diffsync) from the viewpoint of mobile
devices. Diffsync is a synchronisation algorithm which is originally
designed for near real time synchronisation over the web, like 
online computer games or collaborative document editing. We showed 
that is it possible to configure the diffsync algorithm to use push 
notifications and with it retaining the correctness of the algorithm 
and saving energy at the same time still keeping the near realtime 
nature of the algorithm in tact~\cite{Simon2014}. In this paper 
we want to explore the energy characteristics in more detail.
\\
A preliminary version of this paper was published as a poster at the
Mobiquitous 2014~\cite{Simon2014}. This paper extends the poster by an
in-depth analysis of the energy characteristics of different cycle
times, the energy characteristics of size and complexity of single
diffs on the CPU, describe the experiment setup and use case in more
detail in order to give readers the possibility to repeat similar
experiments, and also go more in depth on how the push notifications
work and save energy.

\section{Use Case: Social Reference Management}

Examples in this paper are based on real usage data of editing PDF
annotations within the Mendeley iOS app.
Mendeley is a social document and content reference
management tool, e.g. it enables users to manage their PDF documents,
generate citations and bibliographies. In the Mendeley iOS app, PDF
content can be annotated, e.g., highlighting, or adding and editing
textual notes. Together with other metadata, annotations are stored
locally on the mobile device as well as centrally on the Mendeley
server. References, and their related PDFs can be shared amongst a
group, and PDF annotations can therefore be edited by multiple users.
Data integrity between server and Mendeley app (client) are a
fundamental requirement. This raises the need for synchronising data
between client(s) and server. Since annotations can be done collaboratively,
diffsync was a natural choice for mendeley, contrary to simpler direct
synchronisation schemes or rsync schemes services like dropbox use it. 
Mendeley used the diffsync algorithm in their official iOS client app 
during the time of the study. We can use this data to report realistic 
usage schemes and package sizes for later calculations.




\section{Related Work}

\subsection{Energy Optimisation on Mobile Devices}
Mobile devices such as smartphones and tablets rely on battery
power. Energy intensive hardware and software operations limit the use
by "shortening battery life" and significantly impact user
experience. Energy consumption can be reduced either by modifying
mobile device hardware or mobile applications (software). In this
paper, we are concerned with the latter. For application developers
the challenge is to balance the need for performance
(e.g. computational speed) while maintaining a low level of power
consumption.  For instance, in mobile phone sensing, it is common to
reduce sampling frequency of sensors to reduce energy
consumption. This may lead to reduced precision however~\cite{Ra2012,Rachuri2011}. 
Moreover, the most energy consuming components in mobile
devices are CPU, network components (WiFi, 3G, GSM) and
display~\cite{Carroll2010,Crk2009}\footnote{\url{https://source.android.com/devices/tech/power.html}}. As
optimising display energy consumption is out of application
developers' control except in game development, application developers
are mostly left to optimise CPU and network energy
consumption. In~\cite{Balasubramanian2009}, a network energy
consumption model is described based on empirical data, which
emphasises the possibility to optimise network energy consumption by
exploiting the tail energy property of different network
protocols. This model has influenced other research e.g. for
optimising computational complexity on the mobile device by offloading
computation to the cloud~\cite{Cuervo2010,Gao2013}. Network energy
consumption can sometimes be optimised by choosing suitably between
push and pull communication paradigm, as has been discussed
in~\cite{Burgstahler2013} in connection with the Google cloud
messaging system.  CPU energy consumption can be optimised by adapting
algorithms to mobile devices, as has been done e.g., for the hash
function~\cite{Damasevicius2012} and AdaBoost~\cite{Kadlcek2013}.




\subsection{Delay Tolerant or Prefetch Friendly Synchronisation}

Delay tolerant or prefetch friendly synchronisation methods can
drop the constrain that small updates should be delivered frequently.
As we will see, in a scenario where frequent small updates must be
made, tail energy is the main concern. The best strategy to save 
energy is to collect a bulk of data for transmission for a while, and 
transmit at a time of good connectivity (of fetch a lot respectively).
This minimises the tail energy, as this is a constant added only once
after the bulk transmission. Is also minimises transmission energy, as
this is inverse proportional to the bandwidth available. Thus, using 
time slots of good connectivity is an important part of Mobile Cloud
Computing~\cite{Fangming2013}.
Recent research formulates this problem as a system of applications
putting data into queues for transmission, a discrete time, and a cost
function based on a power model. A Lyapunov drift is then used to
compute when data should be transmitted.
AppATP~\cite{Fangming2015} is a cloud
based middleware managing several applications, and decides for
prefetch friendly and delay tolerant apps in a fixed cycle of $60s$
if a transmission should be made. AppATP archives savings of $30$-
$50\%$ on the synchronisation part of a complete system. 
eTrain~\cite{Tan2015} also uses Lyapunov optimisation, and also
optimises several mobile Applications at once. However, it does
not rely on a own middleware, and piggyback the heartbeat
transmissions applications employing push notifications need to
keep the tunnel open to add transmissions there to reduce
wasted tail energy. It can save $12$-$33\%$.
Similar to both Systems our approach uses a fixed cycle time. Similar
to eTrain reducing the waste of energy caused by tail energy is
a central topic. However, we want to keep the near real time
character of the diffsync. Therefore we employ a different strategy
for tail energy (optimising cycle time/tail energy tradeoff and
push notifications). Also these works look at group of apps while
we look at suggestions for an specific algorithm for an app developer
to implement.

\subsection{Realtime Synchronisation}

For a long time, operational transform (OT)~\cite{Ellis1989} has been
the quasi-standard synchronisation algorithm in use. It is used for
instance in Google products like Google Docs or Google Wave. OT is
based on the notion of expressing edits as operations on the
state of an item (e.g. a document). To avoid loss of changes and
ensure data integrity, OT relies on capturing all user edits for each
item.  Once an edit is lost, different item copies will not converge
towards an agreed version again.  This poses a significant challenge
given today's rich user interfaces and feature sets, including: edit
items via typing, cutting, pasting, undo, redo, drag, drop,
etc.~\cite{Fraser2009}\footnote{See also \url{http://sharejs.org.}}.
\\
There is an "intent" problem with OT~\cite{Sun1998}: It may happen,
that a correct merge of two edits is in conflict with the intent of each individual edit. 
E.g. let us assume that two users edit the sequence ``AB'', one person
wishes to delete ``B'', and the other to insert ``12'' after the
``A''. An intent-preserving merge would lead to ``A12''. Depending
on the sequence of transformation operations, however, OT may lead to
``A1B'', which does not correspond to the intent of either
editor. Woot~\cite{Oster2005} (WithOut OT) tries to resolve this problem
by including not only the exact transformation operation but also
contextual information in one operation. However, Woot does not allow
deleting content. A successor, Logoot~\cite{weiss2008} allows
deletions as well as undos~\cite{Weiss2010}.
\\
Differential Synchronization (diffsync)~\cite{Fraser2009} works on
item states only - i.e. knowledge of edit actions is not required by
the synchronisation algorithm. The key notion of diffsync is that the
synchronisation algorithm compares two states of an item, computes the
differences and necessary changes. While some changes may be lost,
different item copies will always converge. Diffsync is used in the
Mendeley iOS app for synchronising PDF annotations, but also in
MobWrite\footnote{\url{https://code.google.com/p/google-mobwrite/}}
for collaborative writing or CoRED~\cite{Lautamaki2012} for
collaborative programming.  More widespread use of diffsync can be
expected in the future, mainly because of the ease of implementation
with fewer possibilities for programming errors when compared to other
synchronisation algorithms. Firstly, client and server code are nearly
identical. Secondly, synchronisation code does not depend on user
interface code. The latter is highly relevant in collaboration systems
available on a broad range of devices. Thirdly, synchronisation code
is independent of the diff and patch algorithm, which makes
synchronisation code independent of the data structure of an
application. Thus, diffsync delegates the problem of preserving edit
intentions to the diff and patch algorithm. It is therefore ``suitable for any
content for which semantic diff and patch algorithms
exist''~\cite{Fraser2009}.
\\
Causal Trees~\cite{Grishchenko2010} mix features of Woot and diffsync,
but break the elegant independence between synchronisation and
diff/patch algorithm that exists in diffsync.


\section{Differential Synchronization}
\begin{figure}
  \centering
  \includegraphics[width=.45\textwidth]{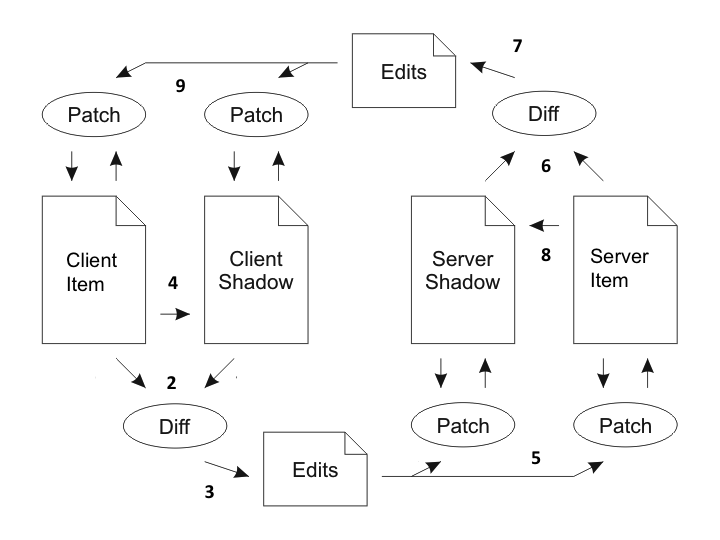}
  \caption{Differential synchronization with shadows - essentially
    corresponds to Fig.3 in \protect\cite{Fraser2009} but
    the numbering and text have been adapted.}
  \label{fig:diffsync}
\end{figure}

Diffsync for one client and one server can be briefly summarised as
follows: Both server and client contain an item (e.g. a document) on
which the diffsync algorithm will be executed. As will be shown below,
both server and client will also create a "shadow" copy of the
item. The initial state is defined as having no edits and assumes
that all copies of the item, including the shadow copies, are
identical. The following list describes one diffsyc cycle starting with an edit on a client item. Figure~\ref{fig:diffsync}
illustrates this procedure.

\begin{enumerate}
\item User edits client item.
\item Diff: Difference between client item and client shadow is computed.
\item A list of necessary edits is the result.
\item\label{it:update-client-shadow} Client item is copied to client shadow.
\item Patch: The edits are patched onto the server shadow (which was identical to the client shadow until Step~\ref{it:update-client-shadow}) and onto
  the server item (which may be different to the server shadow, e.g., in a multi-device/multi-user environment).
\item Diff: Difference between server shadow and server item is computed.
\item A list of necessary edits is the result (edits may be due to
  unsuccessful take-up of client edits, or due to
  edits from other devices/users).
\item Server item is copied to server shadow.
\item Patch: The edits are patched onto the client shadow (after
  applying the patch, the client shadow is identical to the server shadow) and onto
  the client item.
\end{enumerate}

This algorithm can be extrapolated to a system where an item is
shared between multiple clients: The server maintains a central copy of the item (server item)
and a shadow copy for each client (multiple server shadows).

\section{Potential for Energy Optimisation}


There are three areas with potential for energy optimisation. All three
are based on the underlying property of the unmodified differential
synchronisation that the complete diffsync cycle as described above
(Steps 1-9) is executed in regular, fixed intervals.

\subsection{Empty Cycles} The unmodified diffsync executes the
complete diffsync cycles in regular, fixed intervals even if there
are no changes (empty cycles). Clearly, these cycles consume CPU and
network (3G, WiFi, Bluetooth) energy without direct benefit to
synchronisation. 

\subsection{Tail Energy} In 3G and GSM connections, there is a medium
power state lasting for approximately $12s$ for 3G and $6s$ for GSM
once a data connection is terminated.  This tail energy accounts for
up to $60\%$ of the energy consumption of a network
connection~\cite{Balasubramanian2009}. If you transfer at fixed
cycle intervals it is therefore more efficient
to either transfer data more frequently (more than $12s$ resp. $6s$),
or to transfer large amounts of data in less frequent
intervals. Therefore, the worst interval cycle for diffsync would be
$12s$ or $6s$ depending on connection type: The mobile device would
not be able to enter a low power state (network sleep). This leads to
wasting energy in the medium power (tail energy) state, offsetting 
computational advantages by processing very small changes (see next
discussion item). Note that for WLAN connections, there is no such
tail energy~\cite{Balasubramanian2009}, and that different sources
quote slightly different durations of power states\footnote{For
  instance, an additional high power state of $5s$ duration is
  included in the network model at
  \url{http://developer.android.com/training/efficient-downloads/efficient-network-access.html}.}.
Also note that for synchronisation that is delay tolerant it is
more efficient to basically collect transmissions until a large chunk of data
is available and transmit at a time with good connectivity~\cite{Fangming2015, Tan2015, Liu2011}.

\subsection{Computational Complexity} Computational complexity directly correlates
with CPU energy consumption~\cite{Damasevicius2012}. Small changes have a high
likelihood of being computationally less complex than large
changes~\cite{Myers1986}. In addition, item size influences computational complexity, 
as diff is of $O(n^2)$~\cite{Myers1986}.

This discussion leads to the following conclusions: a.) repeated
diffsync cycles containing no changes, i.e. "empty" cycles, should be
avoided. b.) to optimise tail energy consumption, the recommended
cycle interval is below $12s$ for 3G (resp. $6s$ for GSM), or
considerably longer interval durations (than $12s$). Finally c.)
processing small changes and small items reduces computational
complexity and reduces energy consumption. If cycle times can be
considerably longer, Lyapunov optimisation should be considered.

\subsection{Experimental Verification Based on the Mendeley App}

We verified the effects of tail energy and computational complexity on
energy consumption in simulations with the iOS Mendeley App.

\paragraph{Measuring Energy Consumption on Mobile Devices}
Many papers report power consumption of components in $W$ (Watt) and
in $J$ (Joule, $1J = 1Ws$) of events, as these are the correct
physical terms. In order to compute power consumption in terms of $W$
or $J$, system voltage of the battery would need to be measured over
time.  Since we are only interested in the relative energy consumption
of different configurations, we have chosen to measure energy
consumption in terms of percent of battery power. We did this via a
software monitor of battery status, using an iOS internal API which
reports battery drainage in $1\%$ steps. This call requires no
significant energy in case the CPU is already active. A similar setup
to measure energy consumption of apps has been described
in~\cite{oliner-2013-carat} for both Android and iOS. We therefore
express energy consumption in terms of battery drainage in $\%$ per
minute ($\%/min$) in this paper.

\paragraph{Experiment Setup}
We have implemented an experimental framework which includes the
software monitor of battery status and an experiment runner.  The
experiment runner starts executions of the diffsync algorithm as
implemented within the Mendeley app but requires no user
interaction. The runner takes a measurement of battery status before
and after each diffsync cycle. 
Server-side action in the experiments has been simulated based
on real usage data from Mendeley.  With that method more than $60.000$
diffsync cycles were recorded, with about $1.000$ cycles per experiment.
\\
All our experiments run on the same device, an iPhone 5s with a fresh
install of iOS7. The following device setup was used: iCloud was
disabled, one email account with manual updates, no notifications, no
iMessage and no other apps apart from default system apps. In every experiment, unused network
components were switched off in order to avoid battery drainage
through low power states or network scans. The experiment runner dims
the screen to minimal brightness before each experiment run.
\\
The baseline drainage of the environment without execution of diffsync
is $0.108\%$ of battery power per minute. This has been averaged over
a total of 10h recording with sampling time for battery measurement ranging
from $1s$ to $60s$.

\begin{figure}
  \centering
  \includegraphics[width=.45\textwidth]{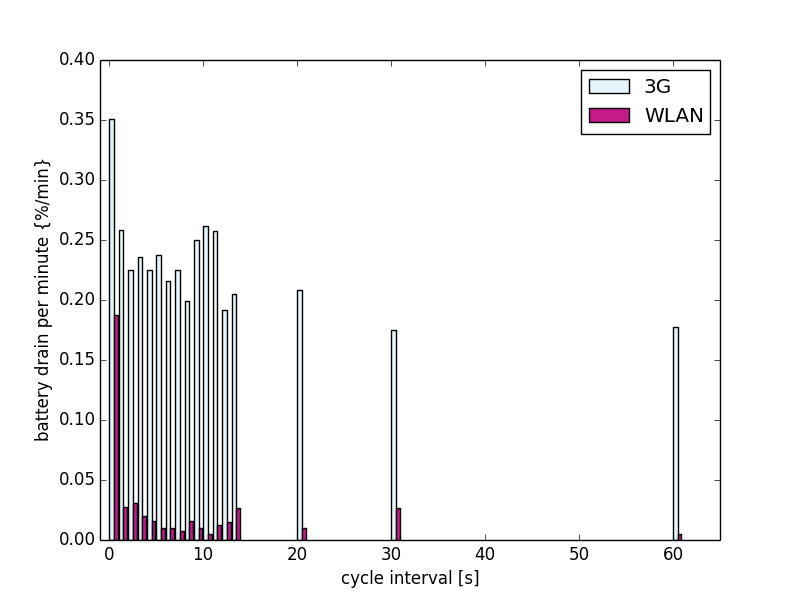}
  \caption{Comparison of battery drainage of 3G vs. WLAN, based on cycle timing}
  \label{fig:3G_WLAN_cycles}
\end{figure}

\paragraph{Tail Energy Consumption}
Based on existing models of network energy
consumption~\cite{Balasubramanian2009}, we have predicted above that
in order to optimise tail energy, the diffsync cycle interval should
either be significantly below $12s$ (3G) or $6s$ (GSM), or
significantly above. For WLAN, we expect no dependance on cycle
interval.
\\
In order to verify this, the experiment runner executed
diffsync with different cycle intervals for 3G, GSM and WLAN network
connection: The cycle intervals were
$0,1,2,3,5,6,7,8,9,10,11,12,13,20,30,60$ seconds. The downlink bandwidths
were 168,2 kbit/s for GSM, 2,4 mbit/s for 3G, and 2,8 mbit/s for WLAN averaged
over all runs.
We observed that a usual package is less than $1kb$ in our case, 
and most of the time about $124bytes$ in diffsync with a $3s$ cycle 
interval. So, for every package size the transfer was always less than
a second, and almost all the energy has to account to cycle time and
cpu complexity. Since we were interested in cycle time energy without
diff energy, and the difference between the standard package ($124bytes$)
and the empty package ($100byte$) are not significant for this test,
we choose to send empty packages around. 
Each experiment was run until $20\%$ of the battery was
drained. The dependance of
diffsync's energy consumption on the cycle interval is illustrated in
Fig.~\ref{fig:3G_WLAN_cycles}.
\\
Experiments show that the predictions based on the network energy
consumption model of~\cite{Balasubramanian2009} are
correct. Additionally, we can experimentally identify a local minimum
around a cycle interval of $6s$ for 3G. In case a fixed and
small cycle interval is desired, $6s$ would then be a recommended
interval for all network protocols: For WLAN, the interval does not
matter, for 3G it is the local minimum, and GSM does not use much less
power even for smaller cycle intervals.



\paragraph{Effect of Computational Complexity on Energy Consumption}
Theory predicts that computationally more complex changes require more
CPU time and, therefore, consume more energy. Computational complexity
for diffsync stems from the complexity of the change, and from the
item size.
\\
We verified and quantified this in an experiment: The experiment
runner executed diffsync with a cycle interval of $6s$. In the first
run, all cycles were empty, i.e. no changes were made. In the second
run, in each cycle a worst case (from a computational complexity
viewpoint) change happened on an item of size $\approx 700 bytes$. In
the third run, in each cycle a worst case change happened on an item
of size $\approx 7000 bytes$. In a fourth run, a ``simple'' change
happened in every cycle on a large item ($\approx 7000 bytes$). In all
four experiment runs, the experiment was run until $20\%$ of the
battery was drained.
\\
When the data structure is an ordered list, the worst case change is a
change in both the beginning and the end of the list for a fixed size
of the change. For instance, if every change has the size of two
characters, it is computationally more complex if ``cat'' is changed
to ``bad'' than if ``cat'' is changed to ``colt''. Ordered lists are a
typical data structure for text, as is the case for PDF text
annotations in Mendeley. A simple change is for instance a deletion at
the end of the item, e.g., ``cat'' is changed to ``ca''.
\\
In the experiment run with empty cycles, energy consumption was
$0.22\%/min$. In the experiment run with the computational worst case
and small item size of $700 bytes$ as well as in the experiment with a
simple change but large item size of $7000 bytes$, no difference in
energy consumption when compared to empty cycles could be measured via
the software monitor. Only for a very large item size of $7000 bytes$
and the worst case change, energy consumption was significantly,
namely $0.34\%/min$.
\\
This confirms the prediction that computational complexity impacts
energy consumption, but only when item size is large and the change is
complex. Since developers can influence item size but not complexity
of changes (this lies with the users), energy consumption due to
computational complexity can best be avoided by reducing item size.
This argues for data structures that break content down into small pieces, so
that diff and patch algorithms can work on small items. For instance,
documents should be broken down into paragraphs or sections.



\paragraph{Generalisation of Experiment Results}
The experiments are performed on iOS, on an iPhone 5s, and on a single
device. Results can nonetheless be generalised to other iPhone 5s
devices, other generations of iPhones, as well as to other platforms
(Android, Windows 8) and phones from different manufacturers: Energy
savings are either specific to the network protocol, like tail energy
to GSM and 3G, or to the CPU. While there are differences in Apple's
Push Notification Service and Google's Cloud Messaging, they rely on
the same principles with respect to network energy
consumption~\cite{Burgstahler2013}. Both platforms use ARM
multiprocessors, and provide similar advice in terms of optimising for CPU
energy efficiency\footnote{For Apple see
  \url{https://developer.apple.com/library/ios/documentation/iphone/conceptual/iphoneosprogrammingguide/PerformanceTuning/PerformanceTuning.html},
  for Android see
  \url{http://developer.android.com/training/monitoring-device-state/index.html}}.
The remaining difference is the operating system: By switching off as
many services as possible in order to still allow the Mendeley app to
be executed, we avoid measuring operating system services such as
background download of emails etc.  Apart from this, the push-based
diffsync optimisation works on an algorithmic level, without
platform-specific elements.

\section{Push-Based Energy Optimisation of Differential
  Synchronisation}

The original diffsync paper~\cite{Fraser2009} suggests adapting cycle
intervals to current editing activities in a client between $1s$ and
$10s$. However, this has been proposed in view of improving
performance (e.g. processing speed, avoidance of merge conflicts),
rather than energy consumption.

In order to strike a balance between computational performance and
power consumption we propose to execute a diffsync only when changes
occur, except for an initialisation cycle. Concretely:

\begin{enumerate}
\item In the initial state, the client connects to the network. In
  order to capture any changes that may have occurred in the meantime
  an initial diffsync cycle will be required.
\item\label{it:edit} When the client item is edited, the client
  initiates a diffsync cycle.
\item If a change arrives at the server, a push notification is sent
  to all clients. On receiving the notification clients execute a
  complete diffsync cycle.
\end{enumerate}

\paragraph{Nearly No Empty Cycles}
Except at initialisation, no empty cycles are being carried out. This
is the central property of the push-based optimisation of energy
efficiency of diffsync. Note that reducing empty cycles only reduces
energy consumption significantly for 3G and GSM connections, as empty
cycles in WLAN drain the battery only minimally.


\paragraph{Correctness} The push-based optimisation algorithm does not
change the diffsync algorithm per se, but only changes the intervals
between cycles. All edits are synchronised to the server and to
connected clients as fast as possible.
\\
Clients not connected to the network (offline) cannot receive push
notifications. In this case clients may have to process significant
amount of changes when reconnecting to the network (online). Depending
on the complexity and nature of changes this may lead to merge
conflicts and failures. However, this is also the case for the original diffsync.

\paragraph{Identifying the Occurrence of an Edit}
The original diffsync algorithm is, amongst other things, easy to
implement because as it is independent of the nature of "an edit''.
The algorithm only needs to compare states regularly.  In contrast,
the push-based optimisation algorithm needs to know what an edit is in
Step~\ref{it:edit}. Note that also Fraser's suggestion for bounded
adaptive timing~\cite{Fraser2009} would require this knowledge.
\\
There are three possibilities how to identify that an edit has
occurred on the client: Firstly, the "diff'' part of the diffsync algorithm could be
executed in regular intervals until a difference (an edit) is
detected. 
Secondly, the data structure (e.g., file, database,
in-memory) in which the application stores its content can be
monitored for changes.  Thirdly, it can be decided which user
interaction means that an edit has occurred. Depending on the
application and which granularity of edits needs to be identified,
this may be as simple as noticing a user pushing a "save'' button, or
as complex as tracking every possible way of editing within a given
UI.
\\
Note that the difficulty (for implementation) of identifying an edit
is lower than in operational transformation (OT). While in the
push-based optimised diffsync, one only needs to identify that an edit
has occurred, OT requires exact knowledge of what the edit consists of
in addition.

\paragraph{Energy Consumption of Push Notifications on Mobile
  Devices}
A network communication overhead is generated via push
notifications. In both Android and iOS, push notifications are
facilitated by a local service (on the mobile device) which regularly
polls for push notifications (Google cloud
messaging\footnote{\url{http://developer.android.com/google/gcm/index.html}},
Apple push
notifications\footnote{\url{https://developer.apple.com/notifications/}}).
\\
We quantified this overhead in an experiment: The local service that
polls for push notifications was run for $2h$ without incoming push
notifications (idle polling). Idle polling adds $\approx 0.02\%/min$
of energy consumption to the idle operating system, thus adding
nearly no overhead. In a $2h$ run with a push notification every $6s$
(active polling), energy consumption was $0.13\%/min$. This overhead
is only for receiving push notifications, not for executing the
diffsync cycles. For instance, if every $6s$ a change on a small item
happens, overall energy consumption would be $\approx
0.35\%/min$. This number stems from adding the energy consumption of
active polling with notifications every $6s$ ($0.13\%/min$) to the
energy consumption of diffcycles every $6s$ for items of small size
($0.22\%/min$ for items of size of $\approx 700 bytes$).
\\
Therefore, the polling required on mobile devices in case of a push
notification mechanism leads to a negligible energy
consumption overhead while no changes occur (idle polling). This means
that the push notification mechanism nearly eliminates the energy
wasted in empty cycles in the original diffsync. However, active
polling introduces an overhead of $\approx 0.13\%/min$. This means
that if changes occur regularly and frequently, it is more energy
efficient to adapt cycle time to editing activity.

\paragraph{Minimum Time Between Push Notifications}
Note that when the interval between push notifications becomes smaller
than $2s$, the mechanism became highly unreliable in our experiments, in the
sense that the sequence of notifications was changed or notifications
became lost. Server code should therefore take care to send
notifications about changes only in intervals larger than $2s$. In
addition, clients need to ensure that only one diffcycle is running at
a time.

\section{Push-Based Diffsync within Mendeley}

We have implemented the above-described push-based energy optimisation
of diffsync in the Mendeley app. In this section, we describe our
implementation, emphasising app specifics in terms of implementation
and usage that impact the actual effect of this optimisation. 

\subsection{Diffsync in the Mendeley App}
In the Mendeley app, diffsync is active in clients when a PDF document
is opened, and inactive when the document is closed. Synchronisation
is on PDF annotations such as textual highlights and notes. The data
content exchanged in each cycle is based on differences in position,
and colour for highlights. When adding and managing notes, differences
in text and author names as well as position and colour will be
added. Therefore, highlighting text results in a smaller diffsync
payload than annotation notes.  The payload also includes other
synchronisation metadata, e.g. unique identifiers.  The format of the
exchanged data set is based on standard JSON.
\\
The Mendeley app's internal data structure for PDF annotations is a
dictionary where every PDF annotation (highlight or note) is a
dictionary entry identified by a unique ID. The ``diff''-part of the
diffsync algorithm therefore works on very small items, namely single
dictionary entries. Thus, computational complexity has no significant
effect on diffsync energy consumption in the Mendeley app. Data are
stored in a Core
Data\footnote{\url{https://developer.apple.com/library/iOS/documentation/Cocoa/Conceptual/CoreData/}}
database.

\paragraph{Original Diffsync}
The diffsync cycle interval in the implementation of the original
diffsync was $2s$ before the push-based optimisation. In empty cycles,
approximately $100 bytes$ of data (metadata) are transmitted.

\paragraph{Push-based Diffsync}
For correctness, we ensure that only one diffsync cycle can run at a
time. If a push notification arrives or a local edit is identified while a diffsync cycle is
running, we let the algorithm perform another cycle right after the
current one. We do so via a boolean flag \texttt{loopOnceMore}. When
diffsync starts a cycle, the flag is set to \texttt{false}. In case a
push notification arrives while a cycle is running, the flag will be
set to \texttt{true}. At the end of each cycle diffsync checks that
flag, and if \texttt{true} performs another cycle.
\\
We identify the occurrence of edits on the client by listening to
modifications of the data structure. Since CoreData sends
notifications about changes to its content in any case, this does not
add extra computation and therefore does not consume additional
energy. 
\\
Servers-side, push notifications are sent to all clients, but waits at
least $2s$ between sending push notifications in order to avoid
unreliable behaviour.

\subsection{Mendeley Usage Statistics}
We analysed Mendeley usage data from a period of $4$ weeks in May
2014. 
In this period, users were far more likely (with an approximate ratio
of $3:1$) to change text highlights than to edit 
textual notes. This strongly supports the conclusion that
computational complexity does not offer potential for energy
optimisation in the Mendeley app, since item sizes are very small not only by design but also by usage.

\paragraph{Average Interaction Time with PDFs}
On average, users spent $431,4s$ ($\approx 7min$) within a PDF,
reading and editing. Of course, the interaction time varies widely,
with a minimum of $3,3s$ and a maximum duration of $1977,8s$ ($\approx
33min$).


\paragraph{Empty Cycles}
A total of $100.000$ diffsync cycles were analysed from the server
logs.
Of these, $96792$ were empty cycles. Thus, an overwhelming majority of
diffsync cycles ($96.8\%$) are empty and waste energy. 

\subsection{Reduction of Energy Consumption via Push-Based Diffsync in
  Mendeley}
$96.8\%$ of diffsync cycles in the non-optimised Mendeley app are
empty, thereby wasting energy.  The push-based optimisation of
diffsync therefore significantly reduces energy consumption. We can
quantify this, using average use data from Mendeley usage
statistics. Assuming, a user stays $7min:$ in a document, and on
average $96.8\%$ of cycles of the original diffsync are empty, and the
user has a 3G connection. The original diffsync with a $2s$ cycle
interval drains the battery $1.729\%$ in these $7min$. The push-based
diffsync however drains the battery only $0.084\%$ in these
$7min$. For the maximum time in a PDF observed in the usage data we
analysed, $33min$, this looks even more drastic: The original diffsync
drains $8,151\%$ of battery power, while the push-based diffsync uses
only $0.398\%$ of battery power.

\section{Conclusion}
Differential synchronisation as realtime synchronisation algorithm for
collaborative editing systems has three potential areas for optimising
energy consumption: Empty cycles, tail energy (cycle intervals) and
computational complexity. We have shown theoretically, and verified in
experiments, that a.) tail energy optimisation argues for cycle
intervals of $\approx 6s$ for 3G and
b.) the impact of computational complexity on energy consumption can
best be addressed by appropriate data structures that organise content
into small items. Tail energy optimisation is useful for instance when
server code cannot be changed, as the cycle interval is determined by
the client.
\\
Most significantly, we have proposed the push-based optimisation of
differential synchronisation, which eliminates empty cycles (except
for initialisation purposes). This optimisation is useful in
collaborative editing systems where edits are rather infrequent, as is
the case in the Mendeley app which we used to showcase the benefits of
the push-based optimisation. In such systems, the push-based
optimisation of differential synchronisation leads to a system with
both higher response time and lower energy consumption.
\\
Overall, we emphasise, that energy optimisations of differential
synchronisation should be done based on knowledge of a collaborative
system's usage data.


\section{Acknowledgments}
The Know-Center is funded within the Austrian COMET Program under the
auspices of the Austrian Ministry of Transport, Innovation and Technology, the
Austrian Ministry of Economics and Labor and by the State of Styria. COMET
is managed by the Austrian Research Promotion Agency FFG.

%
%
%
%
%

\bibliography{diffsync-no-links}{}
\bibliographystyle{agsm}
\end{document}